\documentclass[12pt]{article}
\usepackage{epsfig}
\usepackage{cite}
\usepackage{here}
\input amssym.def
\input amssym

\def\beq{\begin{equation}}
\def\eeq{\end{equation}}
\def\bea{\begin{eqnarray}}
\def\eea{\end{eqnarray}}

\begin{document}

\rightline{May 2006}

\vspace{0.4cm}

\begin{center}
{\Huge\bf On correlations and discreteness}\\
\vskip 0.3cm
{\Huge\bf in non-linear QCD evolution}\\
\vspace{0.8cm}
 
N.~Armesto$^{1}$ and
J.~G.~Milhano$^{2,3}$
 
\vspace{0.2cm}
 
{\it $^1$ Departamento de F\'{\i}sica de Part\'{\i}culas and
IGFAE,\\
Universidade de Santiago de Compostela,
 15782 Santiago de Compostela, Spain}
\vskip 0.1cm
{\it $^2$ CENTRA, Instituto Superior T\'ecnico (IST),\\
Av. Rovisco Pais, P-1049-001 Lisboa, Portugal
}
\vskip 0.1cm
{\it $^3$ Departamento de F\'{\i}sica, FCT, Universidade do Algarve,\\
P-8000-117 Faro, Portugal
}
\end{center}

\vskip 0.3cm
{\small
We consider modifications of the standard non-linear QCD evolution in an
attempt to account for some of the missing ingredients discussed recently,
such as correlations, discreteness in gluon emission and Pomeron loops.  The
evolution is numerically performed using the Balitsky-Kovchegov equation on
individual configurations defined by a given initial value of the saturation
scale, for reduced rapidities $y=(\alpha_s N_c/\pi)\,Y<10$.  We consider the
effects of averaging over configurations as a way to implement correlations,
using three types of Gaussian averaging around a mean saturation scale.
Further, we heuristically mimic discreteness in gluon emission by considering
a modified evolution in which the tails of the gluon distributions are
cut-off.  The approach to scaling and the behavior of the saturation scale
with rapidity in these modified evolutions are studied and compared with the
standard mean-field results.  For the large but finite values of rapidity
explored, no strong quantitative difference in scaling for transverse momenta
around the saturation scale is observed. At larger transverse momenta, the
influence of the modifications in the evolution seems most noticeable in the
first steps of the evolution.  No influence on the rapidity behavior of the
saturation scale due to the averaging procedure is found. In the cut-off
evolution the rapidity evolution of the saturation scale is slowed down and
strongly depends on the value of the cut-off.  Our results stress the need to
go beyond simple modifications of evolution by developing proper theoretical
tools that implement such recently discussed ingredients.  }

\newpage

\section{Introduction} \label{intro}

The
B-JIMWLK\,\footnote{Balitsky--Jalilian-Marian--Iancu--McLerran--Weigert--Leonidov--Kovner.}
equations, derived over the last decade as a result of the concerted effort of
several
groups~\cite{McLerran:1993ni,McLerran:1993ka,McLerran:1994vd,Jalilian-Marian:1996xn,Jalilian-Marian:1997gr,Jalilian-Marian:1997dw,Kovner:1999bj,Kovner:2000pt,Iancu:2001ad,Iancu:2000hn,Ferreiro:2001qy,Balitsky:1995ub,Mueller:2001uk,Blaizot:2002xy,Weigert:2000gi},
build upon the original ideas on gluon saturation set out
in~\cite{Gribov:1984tu} to provide the contemporary  description of  the
evolution of QCD scattering amplitudes at high energy.  Unfortunately, these
are complicated equations and their complete solution is unknown.  A
combination of
numerical~\cite{Braun:2000wr,Kimber:2001nm,Armesto:2001fa,Levin:2001et,Lublinsky:2001bc,Golec-Biernat:2001if,Albacete:2003iq,Albacete:2004gw,Rummukainen:2003ns}
and analytical~\cite{Levin:1999mw,Iancu:2002tr,Mueller:2002zm,
Mueller:2003bz,Munier:2003vc,Munier:2003sj,Munier:2004xu} studies has
established the asymptotic properties of the B-JIMWLK equations.  Most of
these studies have focused on the mean field limit of the equations, where the
B-JIMWLK set reduces to a single closed
equation~\cite{Balitsky:1995ub,Kovchegov:1999yj} - the Balitsky-Kovchegov (BK)
equation.  Nevertheless, the results obtained for the BK equation are known to
deviate at the most $\sim 10$\% from those obtained numerically for the full
B-JIMWLK~\cite{Rummukainen:2003ns}.

In obtaining this set of equations, the gluonic density in the target was
assumed to be large, and the projectile was taken as a dilute object.  The
strict validity of the resulting evolution scheme is, therefore, restricted to
this physical domain.  Further, it has become clear that the physical region
of interest - i.e. that within reach of accelerators that are currently
operating and those that will be operational in the near future - lies within
the pre-asymptotic region of the evolution and that in this region the
evolution is dominated by the, as yet poorly understood, initial
conditions~\cite{Armesto:2004ud,Albacete:2004gw,Albacete:2005ef}. 

In the last two years, there has been a spurt of activity  in this domain
triggered by the crucial observation that important effects for the evolution
are absent from the currently available evolution
scheme~\cite{Levin:2003nc,Mueller:2004se,Iancu:2004es,Iancu:2004iy}.  These
new contributions stem from the understanding of the importance of gluon
fluctuations and have been variably referred  to as 'Pomeron
loops'~\cite{Iancu:2004iy,Iancu:2005nj,Mueller:2005ut,Levin:2005au,Marquet:2005hu,Kovner:2005nq,Kovner:2005en,Kovner:2005uw},
'fluctuations'~\cite{Iancu:2004es} or 'wave function saturation effects'.
A particularly important development has been the finding of a duality
transformation connecting the high and low density
regimes~\cite{Kovner:2005en,Kovner:2005uw}. This duality places strict
constraints on the form of the evolution kernel in the intermediate region.
Nonetheless, the evolution in this intermediate domain is not completely
understood.

Until now all analysis of the effect of these new
ingredients have been addressed, both analytically and numerically, through
the analogy of high-energy QCD evolution to a reaction-diffusion
process~\cite{Iancu:2004es}. These discussions are based on the study
of the stochastic Fisher-Kolmogorov-Petrovsky-Piscounov (sFKPP)
equation~\cite{Iancu:2004iy,Iancu:2005nj,Soyez:2005ha,Enberg:2005cb,Brunet:2005bz,Marquet:2005ak}.
The main conclusions extracted from these studies are: i) at small rapidities,
mean field evolution through the BK equation remains approximately valid, but
the rapidity evolution of the saturation scale is slowed down and the domain
of geometric scaling is restricted, in comparison with the results of pure BK
evolution; ii) at large rapidities, geometric scaling is destroyed, being
replaced by a new type of scaling, sometimes referred to as
diffusive~\cite{Hatta:2006hs}.

At this stage, and while efforts towards the derivation of the complete
evolution take place, it appears possible to attempt to implement some of the
effects as effective modifications of the BK equation, and to examine whether
the results agree qualitatively or not with our current knowledge and
expectations from the results of the, previously mentioned,  full high-energy QCD evolution.  In this paper, we assess the impact of some of the proposed new
contributions.  In particular, we address the effect of correlations by
introducing an averaging procedure for the initial conditions for the
evolution. Further, we model possible effects due to the discreteness in gluon
emission by considering the BK evolution of gluon distributions with cut-off
tails.

The paper is organized as follows. In Section 2 we outline the numerical
method used to implement the BK equation, and the procedures adopted to
introduce modifications in the evolution intended to mimic the effects of
correlations and gluon discreteness. Our results are presented in Section 3
and discussed in Section 4.

\section{Numerical method}
\label{numerical}

Non-linear QCD evolution is implemented using the BK
equation~\cite{Balitsky:1995ub,Kovchegov:1999yj}
in momentum space in the local approximation
(i.e. neglecting the impact parameter dependence). This approximation should
be valid for a large nucleus in the region far from its periphery.  The
evolved object is the unintegrated gluon distribution $\phi(k,b)\equiv
\phi(k)$ with $k$ the transverse momentum, $b$ the impact parameter and where
the dependence on rapidity $Y$ is implicitly understood. The unintegrated
gluon distribution is related to the scattering probability of a $q\bar{q}$
dipole of transverse size $r$ on a hadronic target, $N(r,b)\equiv N(r)$,
through
\beq
\phi(k)=\int {d^2r \over 2 \pi r^2} \,e^{-ik\cdot r} N(r).
\label{eq1}
\eeq
For the azimuthally independent piece of the solution, which gives the
dominant contribution at large rapidities, the BK equation reads
\beq
{\partial \phi(k) \over \partial y} = \int {dk^{\prime 2}\over k^{\prime 2}}
\left[ {k^{\prime 2}\phi(k^\prime) - k^{2}\phi(k) \over
\left|k^{\prime 2}-k^2\right|} + {k^2 \phi(k) \over \sqrt{4k^{\prime 4}+k^4}}
\right]
-\phi(k)^2,
\label{eq2}
\eeq
with the reduced rapidity $y=(\alpha_s N_c/\pi)\,Y$. In this equation
$\alpha_s$ is fixed.  For $\alpha_s=0.2$, the maximum $y=10$ considered in
this paper corresponds to large physical rapidities $Y\sim 50$.

The equation is solved using a 4th-order Runge-Kutta algorithm with step
$\Delta y=0.025$. The integral is evaluated in the domain $-15<\ln{k}<35$
using a Gauss-Chebyshev quadrature with 400 points. Henceforth units of
transverse momenta will be GeV, with the units corresponding to other
quantities to be read from the respective equations.  By varying the
integration limits, the number of points, the step $\Delta y$, and by
comparison  with previous
algorithms~\cite{Braun:2000wr,Armesto:2001fa,Albacete:2003iq}, the accuracy of
the computations can be estimated to be better than 1 \% in a wide region
excluding one order of magnitude from the limits of the domain. It is usually
much better than that.  The size of this domain allows us to safely study the
evolution up to $y=10$.

\subsection{Initial conditions}
\label{initial}

We will use Eq.~(\ref{eq2}) to evolve individual configurations at initial
rapidity $y=0$, characterized by some functional form and by a given value of
the saturation scale $Q_{s0}\equiv Q_{s}(y=0)$. The first functional form
(GBW) essayed is motivated by the Golec-Biernat-W\"usthoff
model~\cite{Golec-Biernat:1998js},
\beq
\phi_{GBW}(k)=-{1\over 2}\ {\rm Ei}\left( {k^2\over Q_{s0}^2}\right),
\label{eq3}
\eeq
with Ei
the exponential-integral function. The second functional form (MF) reads 
\beq
\phi_{MF}(k)=\gamma_E+\Gamma(0,\xi)+\ln{\xi}, \ \ \xi = \left( Q_{s0}^2\over
k^2\right)^\delta,
\label{eq4}
\eeq
with $\gamma_E$ the Euler constant and $\Gamma$ the incomplete gamma function.
This form is motivated by phenomenological studies of geometric scaling in
$\gamma^*$-nucleon and nucleus collisions~\cite{Armesto:2004ud}. It behaves
$\propto \xi$ ($\ln{\xi}$) for $k\gg (\ll) Q_{s0}$. The parameter $\delta$ in
this functional form acts as one minus the anomalous dimension governing the
behavior of $\phi$ for large $k$. We take\,\footnote{Initial values
$\delta>\delta_c$, $\delta_c=0.627\dots$, are known~\cite{Munier:2003vc} to
develop a wave front with $\delta=\delta_c$. On the other hand, initial values
$\delta<\delta_c$ produce wave fronts which preserve this initial value of
$\delta$, a behavior which has been numerically verified
in~\cite{Enberg:2005cb} and apparently contradicts previous numerical
studies~\cite{Albacete:2004gw}. We have checked  that the reason for this
apparent contradiction lies in the different regions of transverse momentum
studied in both references.  The wave front is characterized by the initial
value of $\delta$ for very large values of $k$ studied in~\cite{Enberg:2005cb}
but not considered in~\cite{Albacete:2004gw}.} $\delta=1$.

Values of the saturation scale in the region $10^{-2}\leq Q_{s0}^2 \leq 10^2$
have been explored for both initial conditions.  This region defines our
averaging domain and has been sampled in 284 points which are roughly
equidistant in logarithmic scale.

\subsection{Averaging procedure and cut-off}
\label{averaging}

After evolving up to a given rapidity $y$, we compute the average of $\phi(k)$
as proposed in~\cite{Kovner:2005jc} i.e.
\beq
\langle \phi(k) \rangle_y ={\int_{10^{-2}}^{10^2} dQ_{s0}^2 W(Q_{s0}^2)
\phi(k) \over \int_{10^{-2}}^{10^2} dQ_{s0}^2 W(Q_{s0}^2)}
\label{eq5}
\eeq
is performed.  Two different averaging procedures have been considered: the
linear averaging~\cite{Kovner:2005jc}
\beq
W_1(Q_{s0}^2)=\exp{\left[ -{(Q_{s0}^2-\langle Q_{s0}^2\rangle)^2\over \Delta}
\right]}
\label{eq6}
\eeq
and a logarithmic averaging procedure given by
\beq
W_2(Q_{s0}^2)={1\over Q_{s0}^2}\,\exp{\left[ -{(\ln{Q_{s0}^2}-\ln
\langle Q_{s0}^2\rangle)^2\over \Delta_l}
\right]},
\label{eq7}
\eeq
with $\langle Q_{s0}^2\rangle=1$. Values of $\Delta,\Delta_l=0.01$, 0.1, 1, 10
and 100 have been explored. Let us note that for both for the linear and the
logarithmic case, $\Delta,\Delta_l=0.01$ is almost equivalent to a
$\delta$-function, while for $\Delta,\Delta_l=100$ roughly 50 \% of the
normalization of the function is contained in the region we study,
$10^{-2}\leq Q_{s0}^2 \leq 10^2$. Finally, $W_2$ with $\Delta_l=100$ is essentially 
a flat function, with a variation $\sim 20$ \% in the considered region - thus
it is our model case for a wide averaging. Let us also
indicate that although the
practical relevance of such large values of $\Delta,\Delta_l$ may be
questionable,
our aim in this paper is to examine whether such a  modification of
standard BK evolution has any effect or not - thus our attitude of taking
extreme values of the parameters characterizing the width of the weight
functions for the averaging.

Finally, motivated by the results obtained in the
framework of the sFKPP
equation~\cite{Iancu:2004es,Iancu:2004iy,Iancu:2005nj,Soyez:2005ha,Enberg:2005cb,Brunet:2005bz,Marquet:2005ak}
which indicate that at large rapidities the dispersion of a set of initial
conditions becomes linearly dependent on the rapidity, we will examine the
effects of a logarithmic averaging procedure $W_2(Q_{s0}^2)$ but with a rapidity dependent value
of the dispersion $\Delta_l=y$. Again, the choice of this value is motivated
not by phenomenological considerations but by the requirements of visibility
of the proposed modification in our limited space of sampled initial
configurations, see below.

On the other hand,  and following the suggestion about the discreteness of the
evolution in~\cite{Iancu:2004es}, see also~\cite{Soyez:2005ha,Enberg:2005cb},
we have introduced what we call a cut-off evolution: at every step in the
evolution in Eq.~(\ref{eq2}), values of $\phi(k)<\kappa$ in the r.h.s. of this
equation have been set to 0:
\bea
{\partial \phi(k) \over \partial y} &=& \int {dk^{\prime 2}\over k^{\prime 2}}
\left[ {k^{\prime 2}\phi(k^\prime) \theta[\phi(k^\prime)-\kappa]-
k^{2}\phi(k)\theta[\phi(k)-\kappa] \over
\left|k^{\prime 2}-k^2\right|} + {k^2 \phi(k)\theta[\phi(k)-\kappa]
\over \sqrt{4k^{\prime 4}+k^4}}
\right]\nonumber \\
&-&\phi(k)^2\theta[\phi(k)-\kappa],
\label{eq8}
\eea
with $\theta[x]$ the step function.  Values of the cut-off $\kappa=0.002$,
0.01 and 0.05 have been used, for evolution starting from a given initial
condition with $Q_{s0}=\langle Q_{s0}\rangle=1$. In principle, such cut-off is
proportional to the inverse of the coupling constant, but with an unknown
proportionality constant which prevents us from making such connection with
the value of $\alpha_s$ in the definition of $y$.

\section{Results} \label{results}

\begin{figure}[!ht]
\begin{center}
\epsfig{file=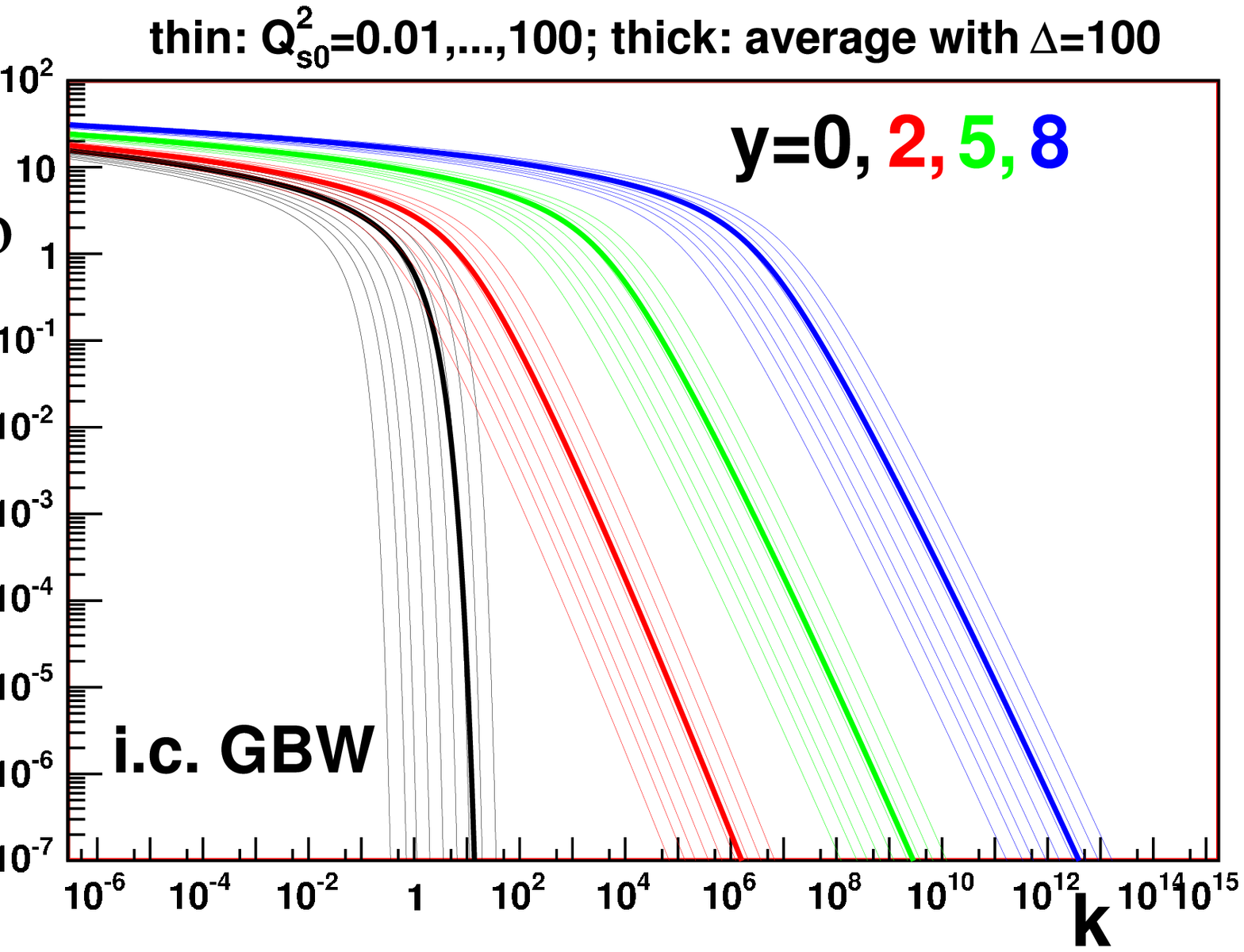,height=7.5cm,angle=90}\hskip 0.5cm\epsfig{file=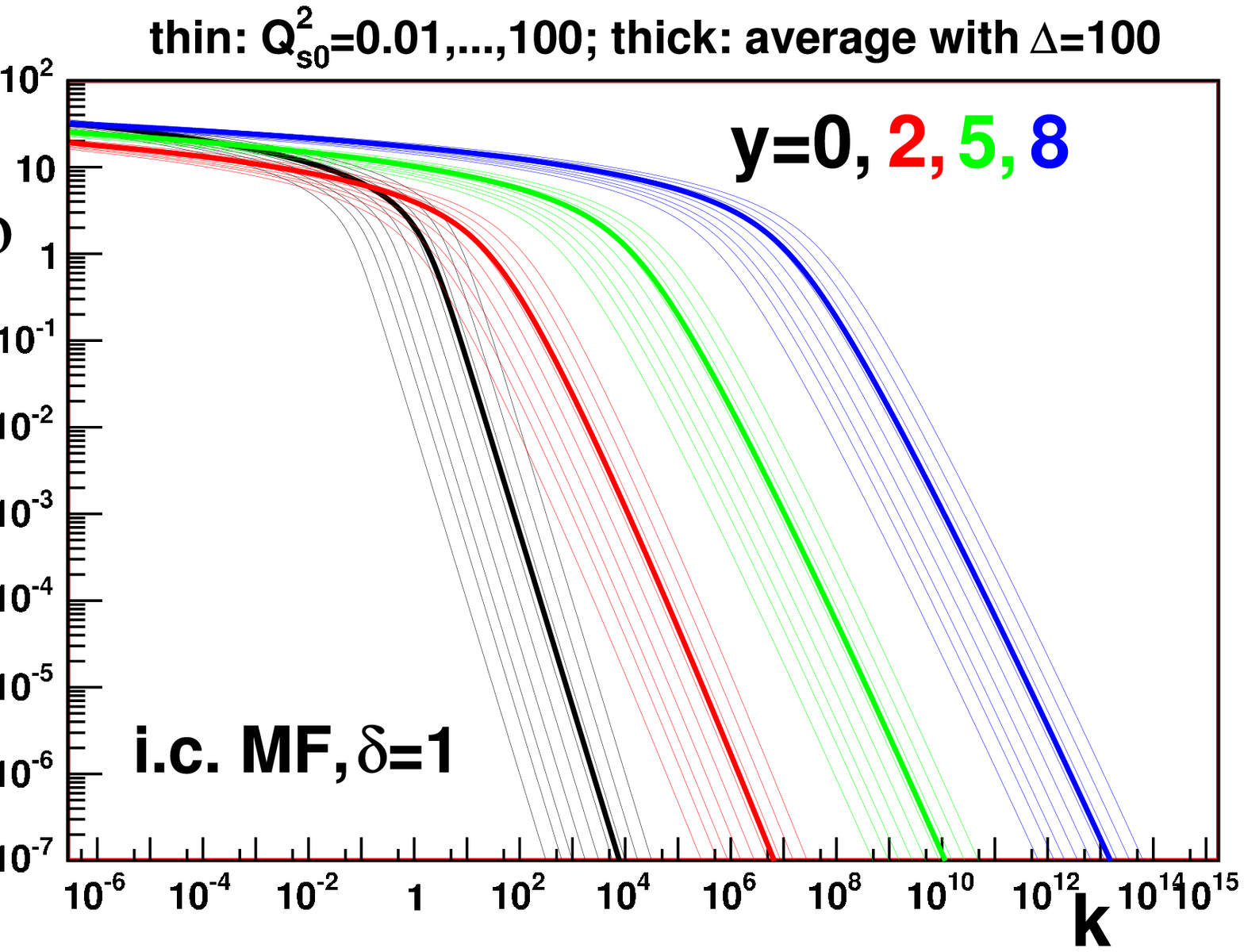,height=7.5cm,angle=90}
\vskip 0.3cm
\epsfig{file=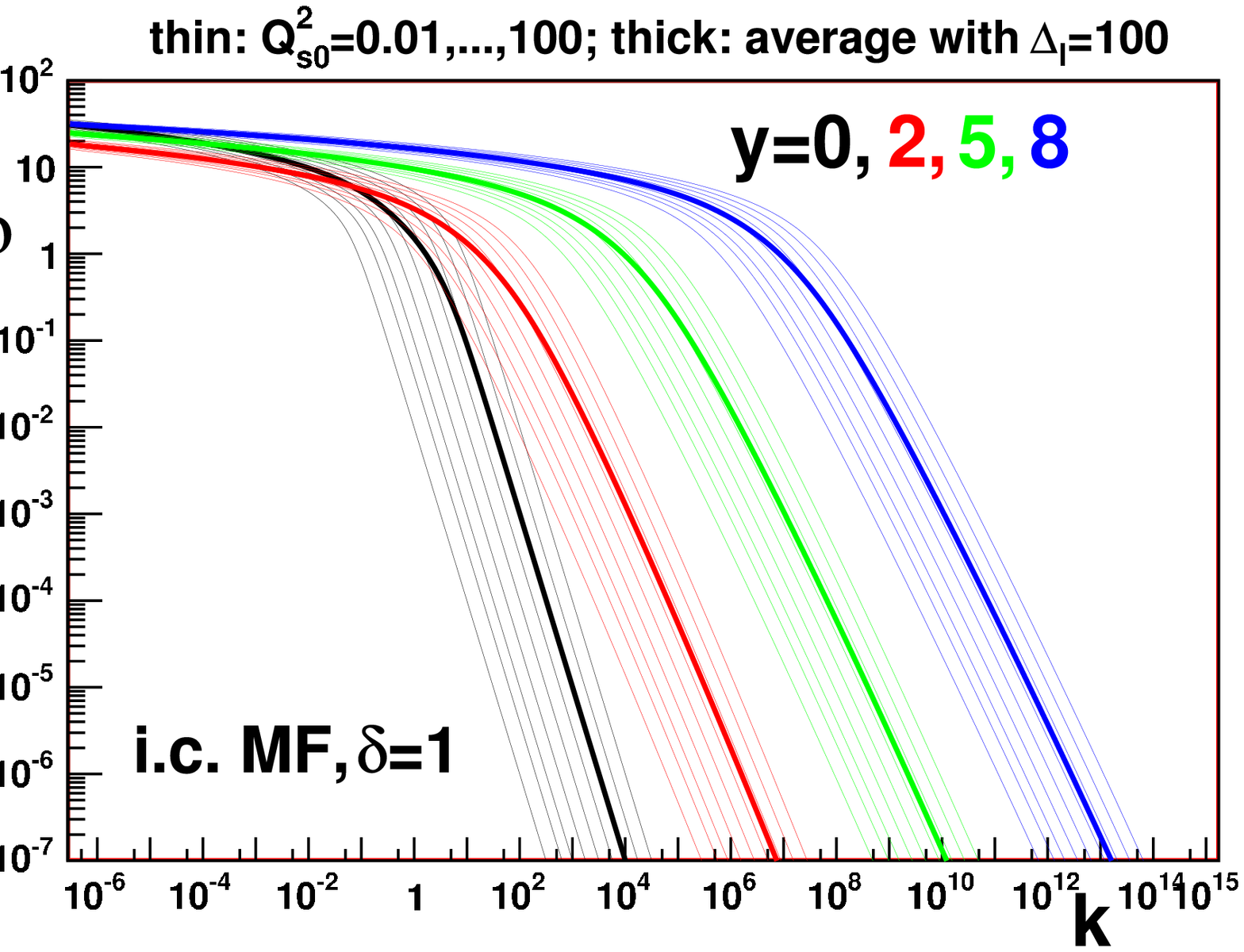,height=7.5cm,angle=90}\hskip 0.5cm\epsfig{file=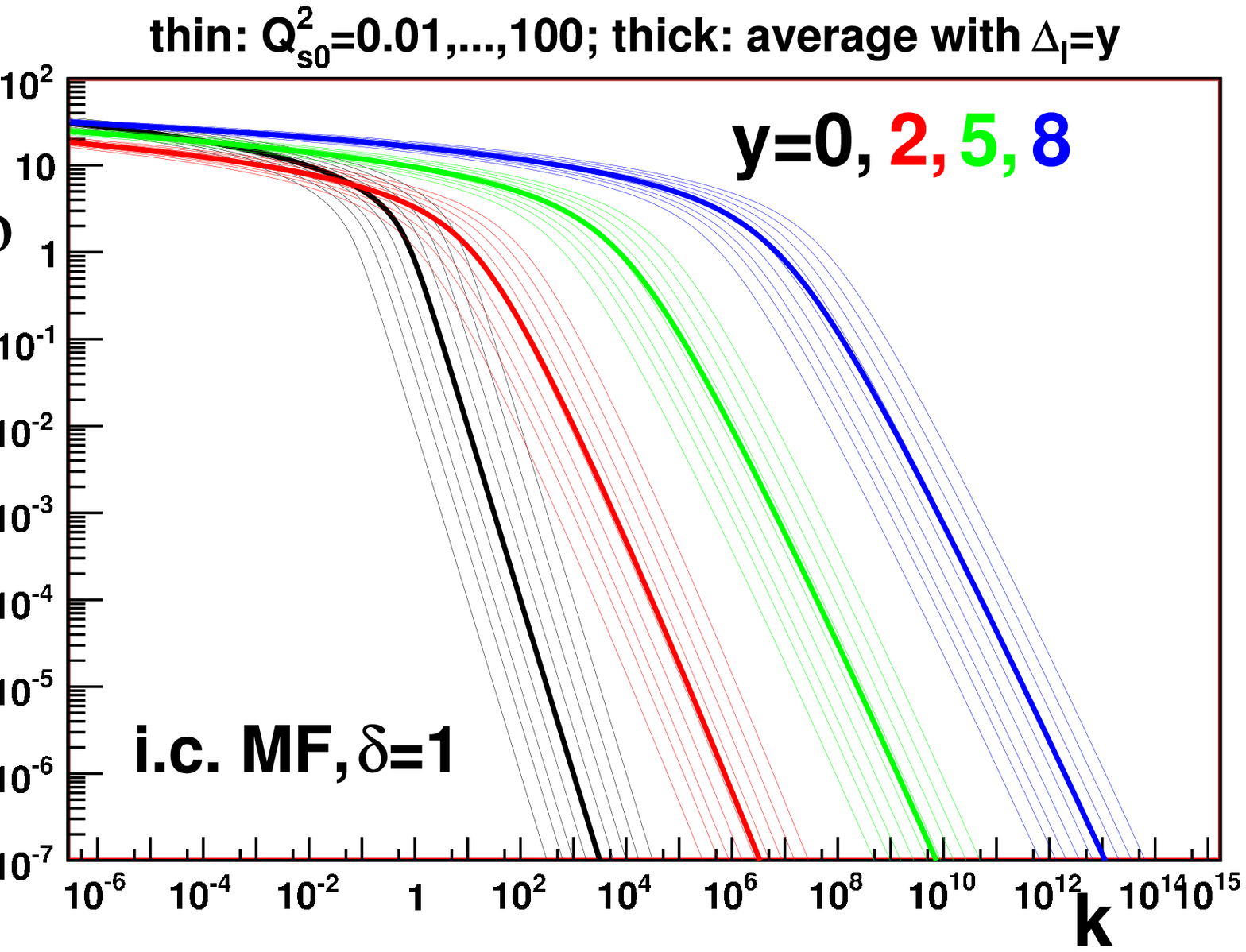,height=7.5cm,angle=90}
\vskip 0.3cm
\epsfig{file=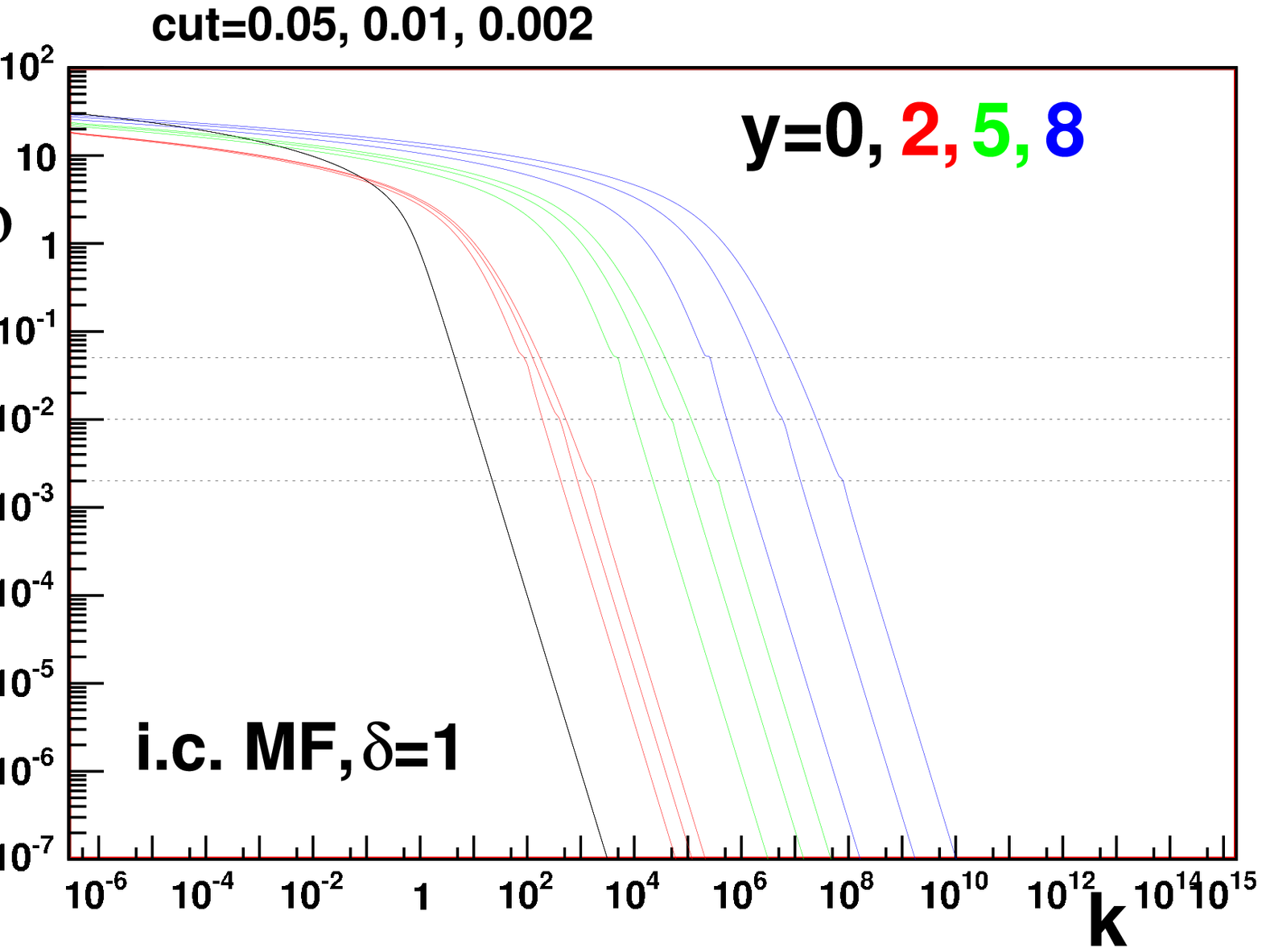,height=7.5cm,angle=90}
\end{center}
\vskip -1.cm
\caption{BK evolution starting from GBW (upper-left plot) and MF (the other
four plots) initial conditions, all for individual configurations with
$Q_{s0}^2=0.01$, 0.04, 0.1, 0.31, 1, 3.31, 10, 30 and 100 (thin
lines left to right),
for linear ($\Delta=100$,
upper plots) and logarithmic averages ($\Delta_l=100$, middle-left plot, and
$\Delta_l=y$, middle-right plot)
using thick lines, and
for cut-off evolution (thin lines, lower plot) for
$\kappa=0.05$, 0.01 and 0.002 (left to right).
In this last plot, horizontal dotted lines
indicate the values of $\kappa$. Results
are shown for $y=0$, 2, 5 and 8 in
black, red, green and blue respectively (sets of lines from left to right).}
\label{fig1}
\end{figure}

We have performed the evolution (\ref{eq2}) starting from the initial
conditions indicated in Subsection~\ref{initial}. In Fig.~\ref{fig1} the
evolution for both initial conditions GBW and MF, for 9 given individual
configurations with $Q_{s0}^2=0.01$, 0.04, 0.1, 0.31, 1, 3.31, 10, 30 and 100,
for linear and logarithmic averages, and for the cut-off evolution\footnote{In
the curves which result from this cut-off evolution, a discontinuity in the
derivative induced by the cut-off is clearly visible.}, are shown. The
development of traveling waves~\cite{Munier:2003vc} can be seen in all cases
and for both initial conditions. For this reason, hereon  we will present
results only for MF, the results for GBW being in agreement.

\clearpage

\begin{figure}[!ht]
\begin{center}
\epsfig{file=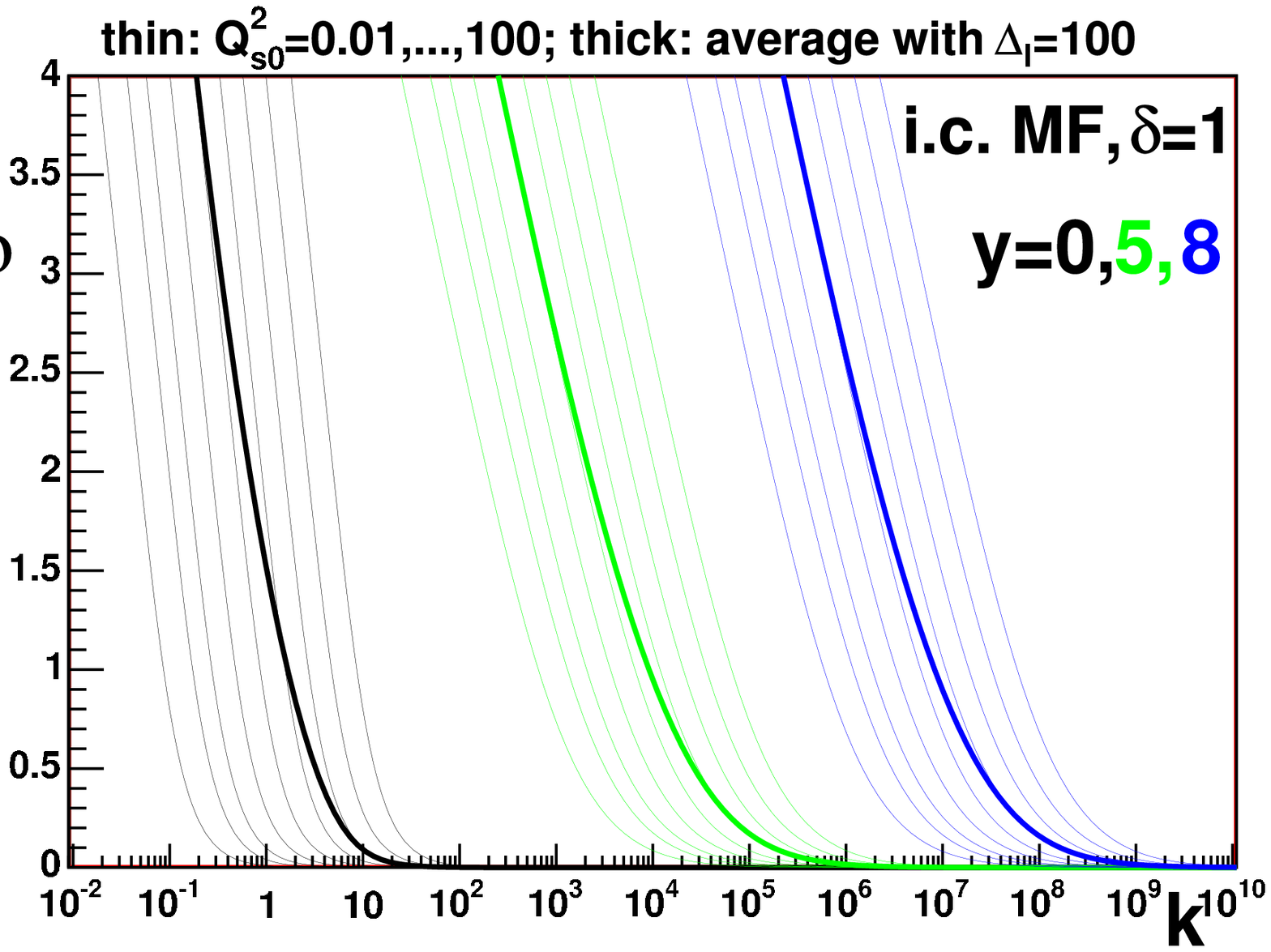,height=14.7cm,width=9cm,angle=90}
\vskip 0.3cm
\epsfig{file=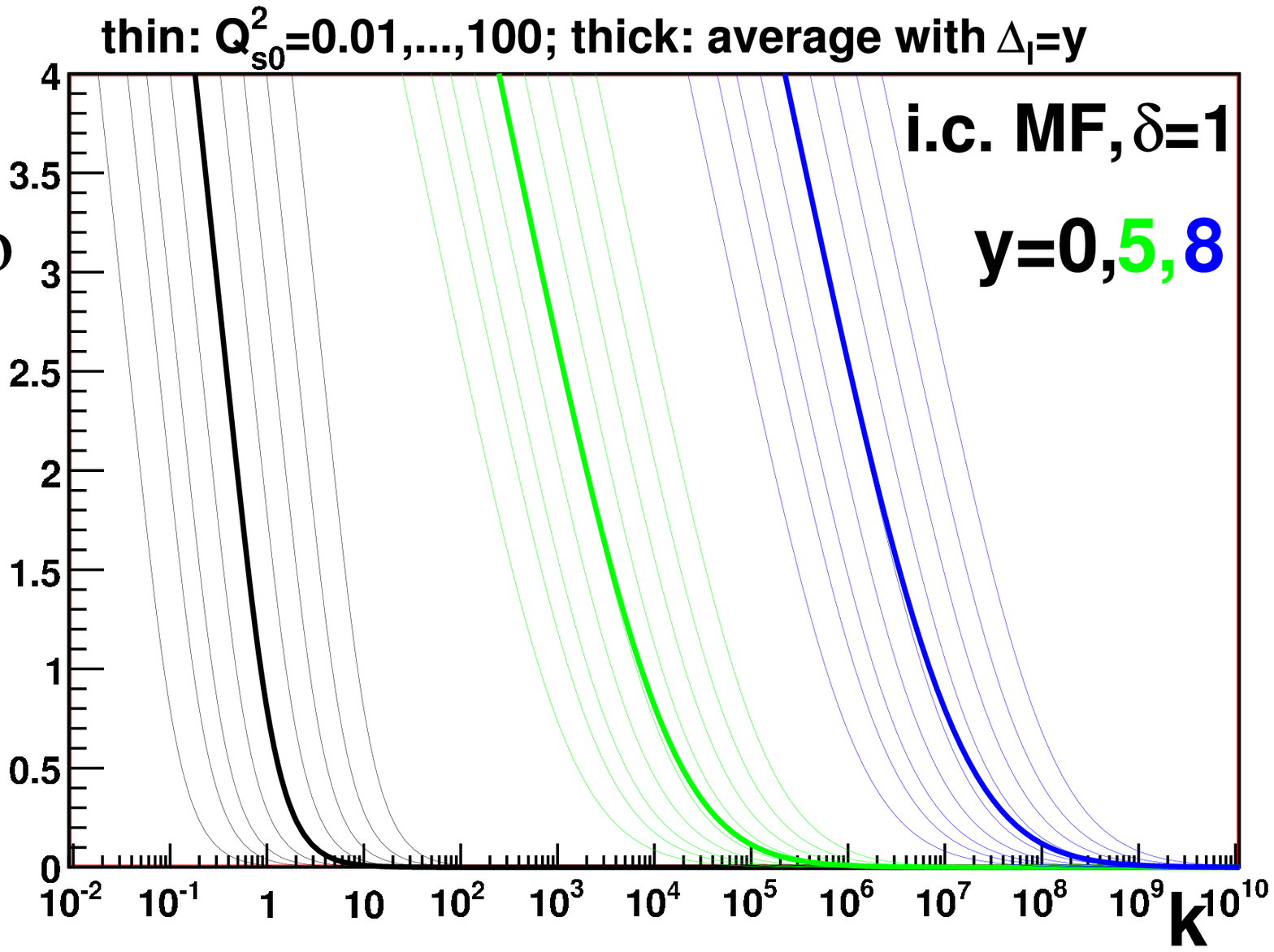,height=14.7cm,width=9cm,angle=90}
\end{center}
\vskip -1.cm
\caption{Details of Fig.~\ref{fig1} (middle-left, upper plot here, and
middle-right, lower plot here) in vertical linear scale.
Results for $y=2$ overlap with those for $y=0$
and are not shown for reasons of clarity.}
\label{fig2}
\end{figure}

In these plots no dramatic effect of the averaging can be observed, the
large-$k$ behavior appearing very similar to that of the individual
configurations. This fact could be expected from the power-like behavior of
the solutions of
BK~\cite{Iancu:2002tr,Mueller:2002zm,Munier:2003vc,Albacete:2003iq} at large
$k$.  In order to see the details of the effect of the averaging we show in
Fig.~\ref{fig2} zoomed versions of Fig.~\ref{fig1} in linear vertical scale. The
pictures clearly show  the mixing of individual configurations attained  by the
averaging procedure, substantiating for BK evolution the picture
in~\cite{Iancu:2004es}.  It can also be seen that the effect of the averaging
with a fixed dispersion is more evident in the initial condition and becomes
less evident with increasing rapidities, in disagreement with present
expectations from the sFKPP equation as commented previously (see
e.g.~\cite{Hatta:2006hs}), while for a dispersion linearly dependent on $y$,
the effect becomes, in agreement with sFKPP predictions, more and more evident with increasing rapidities.

In the next two Subsections we will examine in more detail two features of
evolution, namely the approach to a universal scaling form and the evolution
of the saturation scale with rapidity.

\subsection{Approach to scaling} \label{scaling}

BK evolution is known~\cite{Armesto:2001fa,Lublinsky:2001bc,Albacete:2003iq,
Iancu:2002tr,Munier:2003vc} to approach asymptotically a scaling function,
$\phi(y,k)\to \phi(k/Q_s(y))$ for $y \to \infty$, independent of the initial
condition for evolution. On the other hand,
attempts~\cite{Mueller:2004se,Iancu:2004es} to go beyond the mean field
approximation underlying the BK equation predict a violation of the scaling
behavior, leading to what has been called diffusive
scaling~\cite{Hatta:2006hs}.  Here we try to explore this aspect within our
framework of modified BK evolution.  For the function $\phi$ or the average of
individual configurations, obtained by evolution up to a given $y$, we compute
the saturation scale as the position of the maximum of the function
$h(k)=k^2\nabla_k^2\phi(k)$, in line with  previous
works~\cite{Braun:2000wr,Armesto:2001fa,Albacete:2003iq}\footnote{Note that
this choice corresponds to a region where the function is large, $\phi \sim
1$, and not to a dilute region of the traveling wave. It is
known~\cite{Golec-Biernat:2004sx,Albacete:2004gw,Enberg:2005cb} that with this
definition, sub-leading terms~\cite{Munier:2004xu} in the $y$-evolution of
$Q_s$ are less noticeable. In any
case, we do not attempt to study in detail the behavior of the saturation
scale with rapidity but only to see whether clear differences among the
different modifications of BK evolution appear or not.}. In standard BK
evolution, this function shows a Gaussian-like shape which is preserved in the
evolution, and the velocity of the evolution with $y$ of this soliton-like
form is given by the saturation scale.

\begin{figure}[!ht]
\begin{center}
\epsfig{file=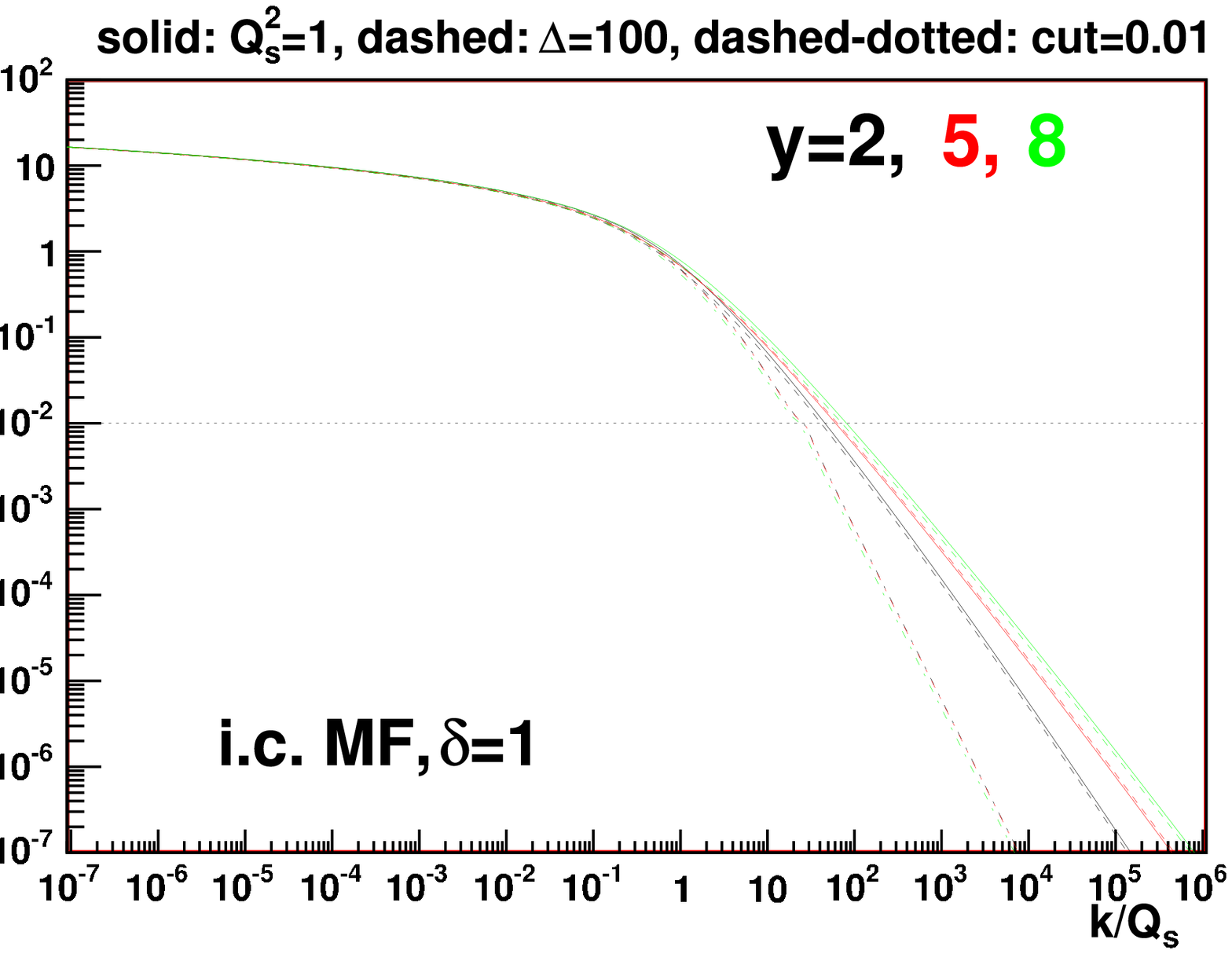,height=7.5cm,angle=90}\hskip 0.5cm\epsfig{file=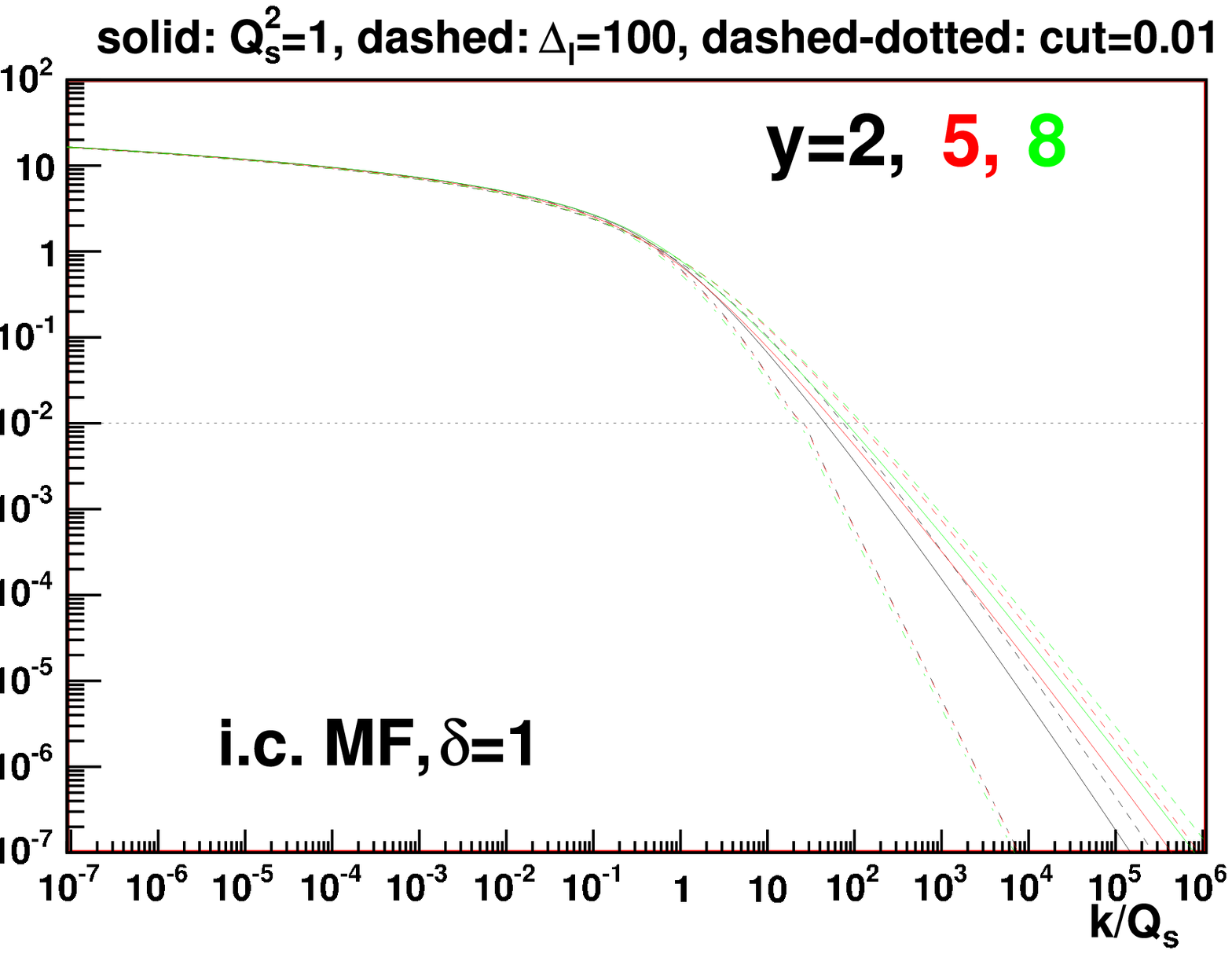,height=7.5cm,angle=90}
\vskip 0.3cm
\epsfig{file=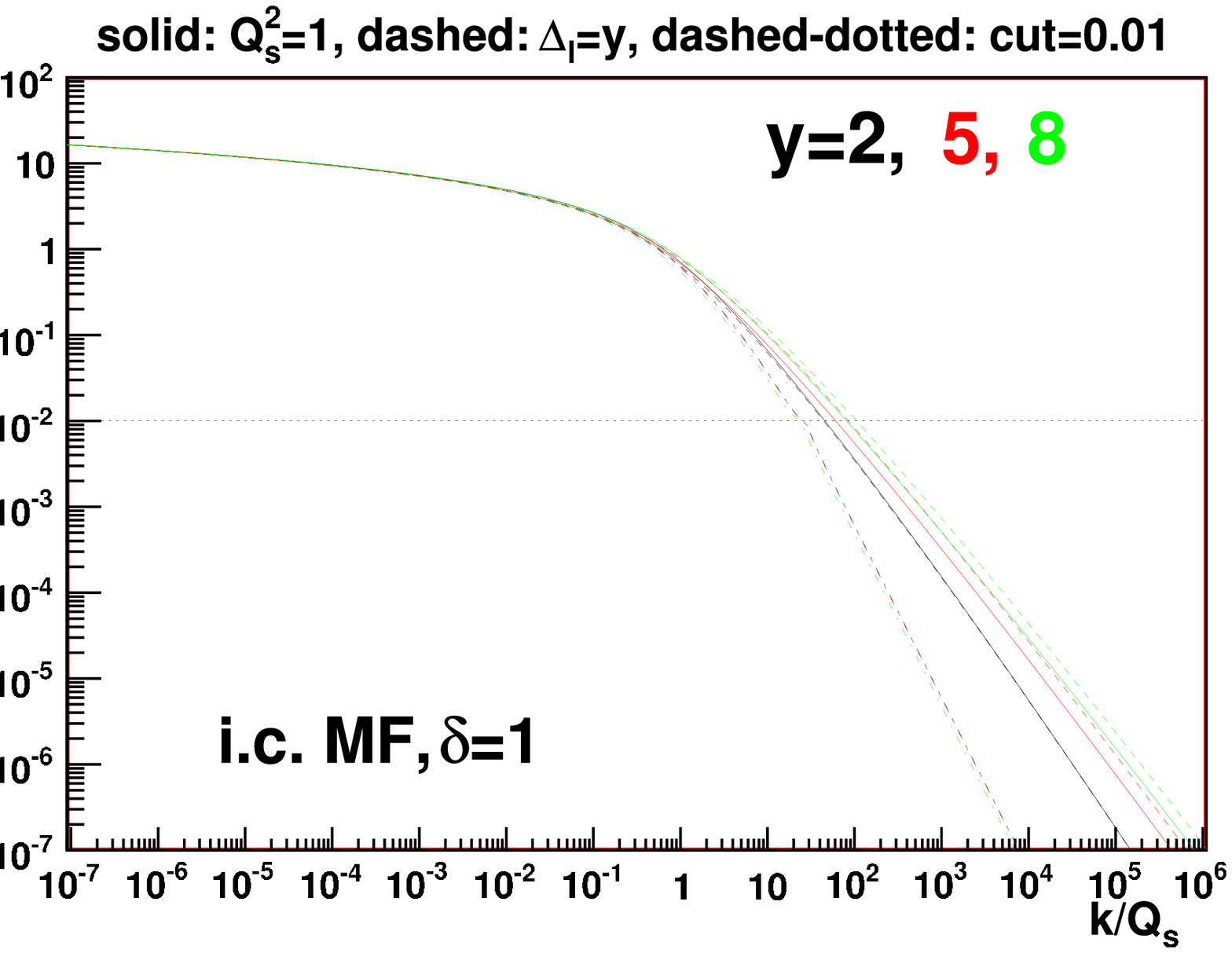,height=7.5cm,angle=90}\hskip 0.5cm\epsfig{file=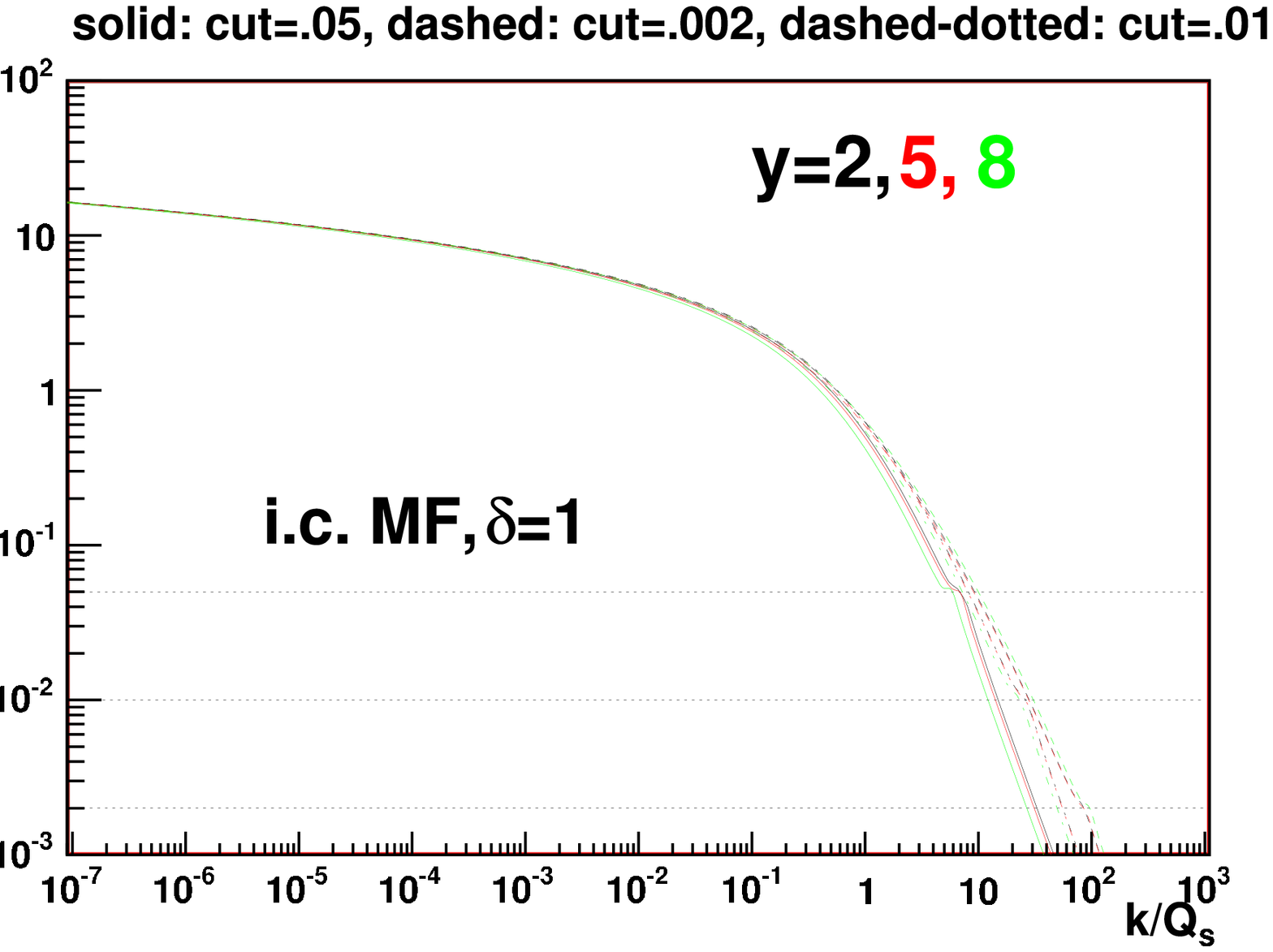,height=7.5cm,angle=90}
\end{center}
\vskip -1.cm
\caption{Scaling forms $\phi(k/Q_s(y))$ for evolution starting from
MF initial conditions, for $y=2$, 5 and 8 (black, red and green lines
respectively). Upper-left plot: results of an individual configuration with
$Q_{s0}=1$ (solid lines), of
linear ($\Delta=100$, dashed lines)
averaging and with cut-off ($\kappa=0.01$, dashed-dotted lines).
Upper-right plot: idem but for logarithmic
($\Delta_l=100$)
averaging.
Lower-left plot: idem but for logarithmic
($\Delta_l=y$)
averaging.
Lower-right plot: comparison of the scaling forms for different cut-offs
$\kappa=0.05$ (solid lines), 0.01 (dashed lines) and 0.002 (dashed-dotted
lines). In this last plot, horizontal dotted lines
indicate the values of $\kappa$.}
\label{fig3}
\end{figure}

In Fig.~\ref{fig3} we show the results for evolution of function $\phi$ when
plotted versus $k/Q_s(y)$.  Curves for an individual configuration with
$Q_{s0}=1$, and for linear and logarithmic averages and for cut-off evolution,
are shown.  The violation of scaling due to averaging does not look dramatic,
the scaling curves being relatively close and getting closer for increasing
rapidities. Only the very wide logarithmic averages seem to induce some
visible violation of scaling, see below. On the other hand, scaling is
achieved very quickly in the cut-off evolution (only for the largest cut-off
and rapidity some departure is visible). This fact could be expected as it is
known~\cite{Iancu:2002tr,Munier:2003vc,Albacete:2004gw} that violations of
scaling come from the tails of large $k$, which are cut in the cut-off
evolution.

\begin{figure}[!ht]
\begin{center}
\epsfig{file=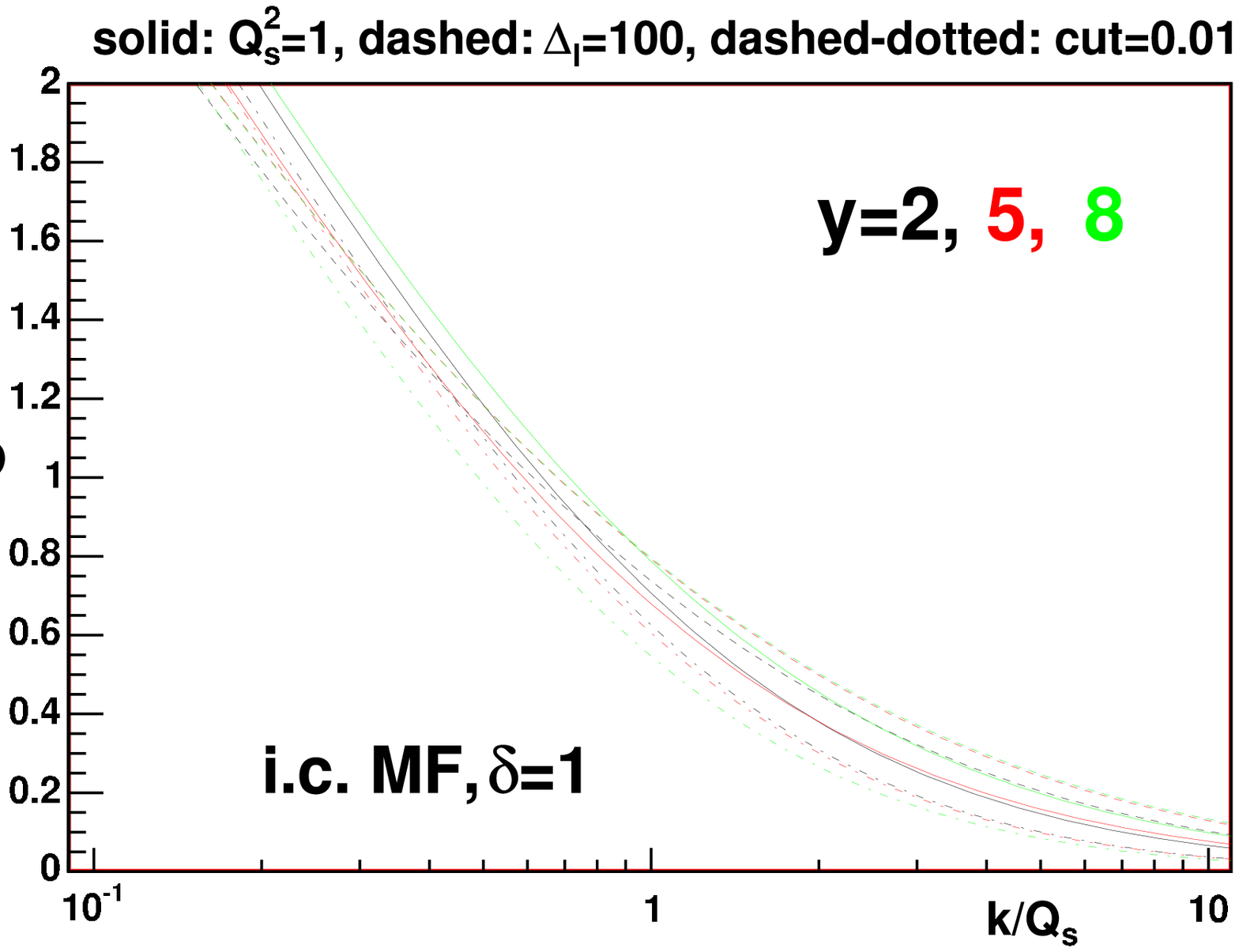,height=14.7cm,width=9cm,angle=90}
\vskip 0.3cm
\epsfig{file=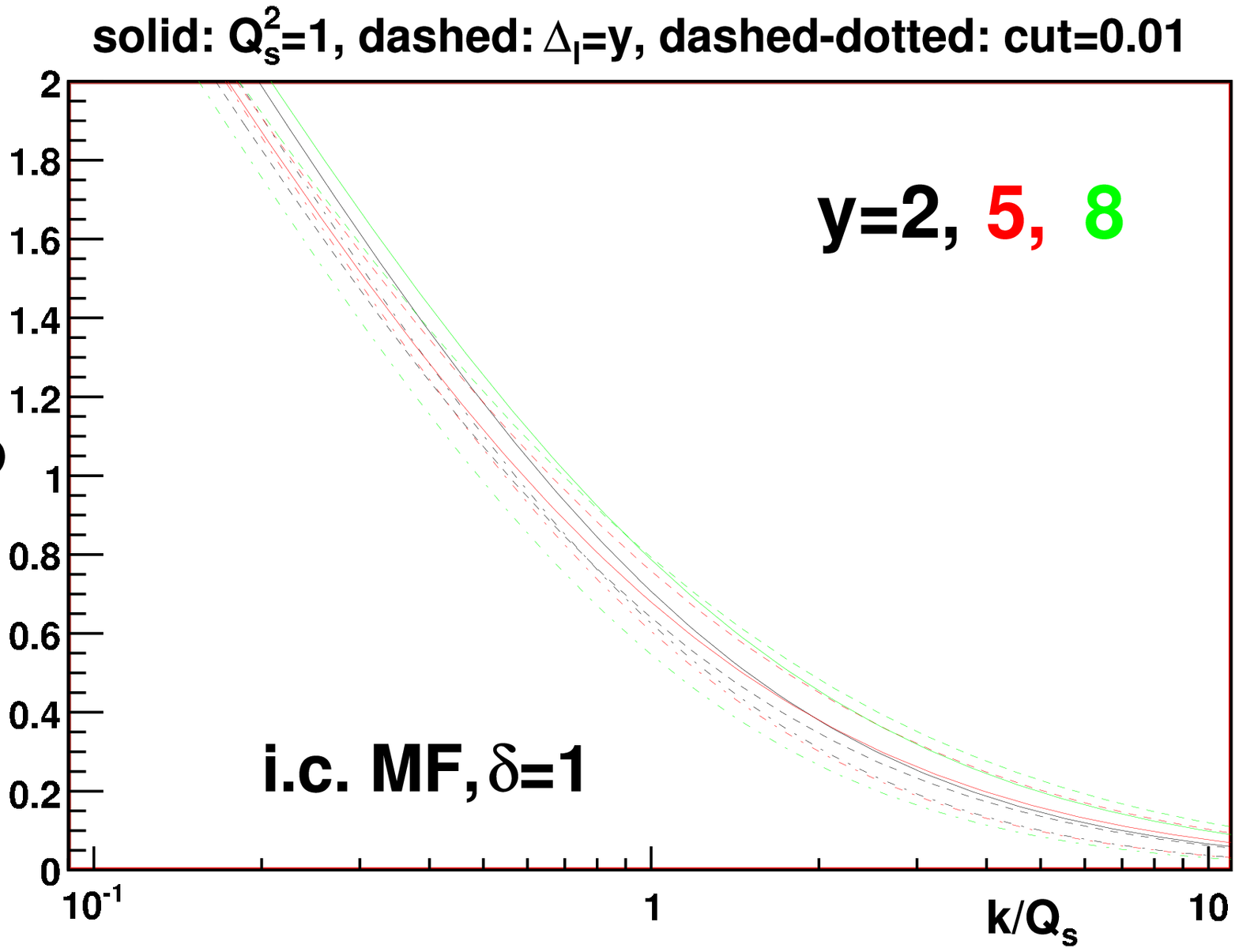,height=14.7cm,width=9cm,angle=90}
\end{center}
\vskip -1.0cm
\caption{Details of Fig.~\ref{fig3} (upper-right, upper plot here, and
lower-left, lower plot here)
in vertical linear scale, for values
of $k$ around $Q_s(y)$.}
\label{fig4}
\end{figure}

To analyze in more detail the effect of averaging on scaling, in
Fig.~\ref{fig4} zoomed versions in linear scale  of the Fig.~\ref{fig3} for the logarithmic averages are
shown. It is well known from standard BK
evolution~\cite{Armesto:2001fa,Lublinsky:2001bc,Albacete:2003iq,
Iancu:2002tr,Munier:2003vc} that individual configurations tend towards a
universal scaling shape for the largest $y$, but violations of scaling of
order $\sim 10$ \% around $Q_s$ persist until large rapidities $y\sim
10$~\cite{Albacete:2003iq}.  The cut-off evolution shows some scaling
violation, smaller than the standard BK results, and which tends to disappear with
increasing $k/Q_s(y)$, see comments above.  Besides, the averaged solutions
for $\Delta_l=100$ show violations of scaling similar those of the individual
solutions at the same $y$. But a dispersion increasing with rapidity,
$\Delta_l=y$, produces larger violations of scaling than the case of standard
BK evolution\footnote{From the comparison of the curves for rapidities
$y=2,5,8$ the violation of scaling induced by the logarithmic averaging with
$\Delta_l=y$ looks $\sim 20$ \% for $k/Q_s(y)=1$, increasing to $\sim 50$ \%
for $k/Q_s(y)=10$, while the corresponding scaling violations for the
individual solutions are smaller.}.

At first sight it looks odd that a
smaller, varying dispersion creates larger violations of scaling than a
larger, fixed dispersion. This is due to the fact that we sample a limited
space of initial conditions, which for a very wide averaging gets immediately
saturated, thus making the effect of the averaging less
noticeable\footnote{This fact justifies our choice $\Delta_l=y$.}. So we
conclude that the effect on scaling of our cut-off evolution seems to
contradict the results from sFKPP evolution (it should be noted that our
cut-off is applied to an evolution performed on averages, not to the evolution
of individual configurations), while the effect on scaling of our averaging
procedure with a dispersion increasing with rapidity
agrees qualitatively with the results from sFKPP.

\subsection{Rapidity evolution of the saturation scale} \label{qsat}

\begin{figure}[!ht]
\begin{center}
\epsfig{file=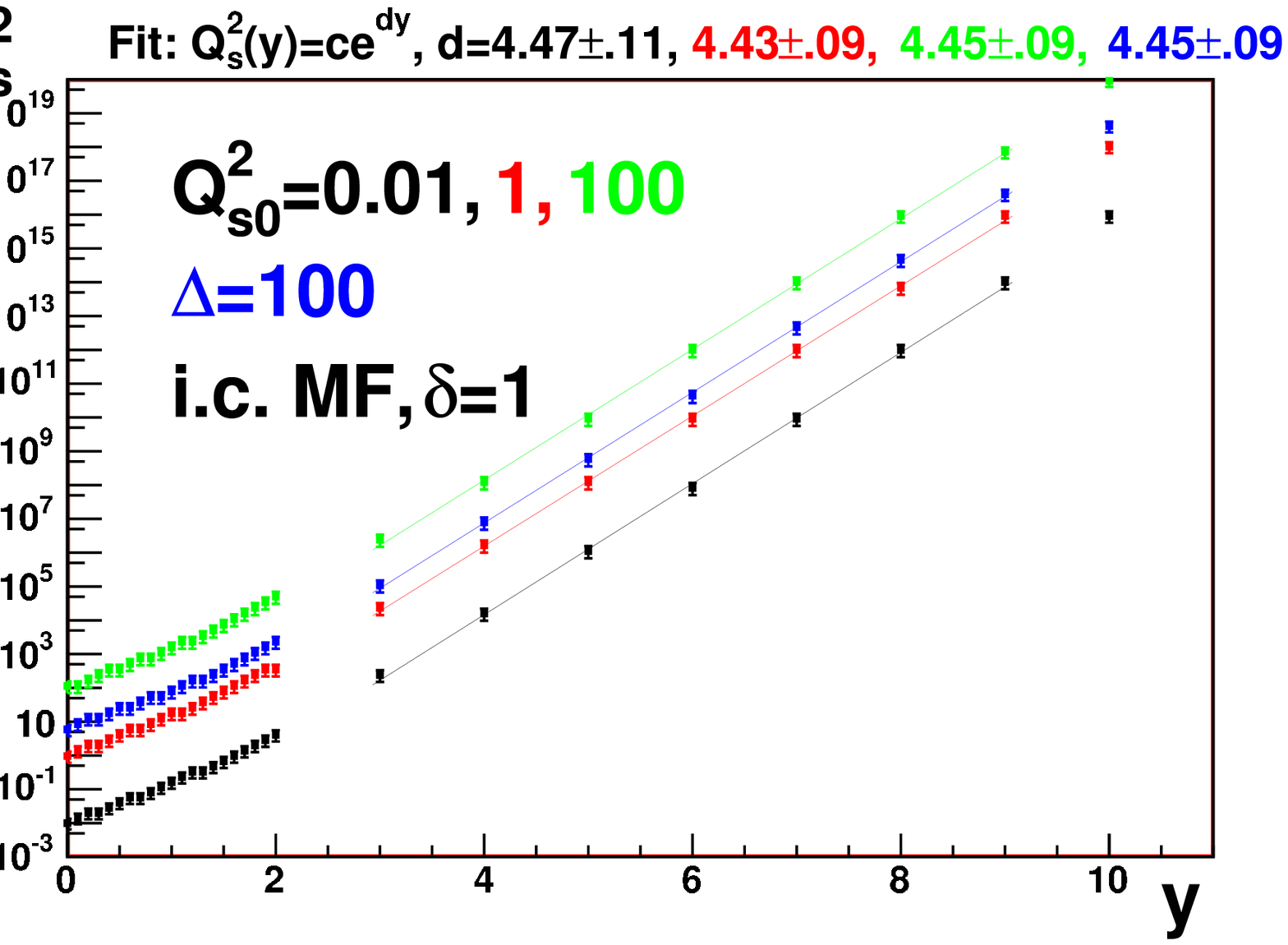,height=7.5cm,angle=90}\hskip 0.5cm\epsfig{file=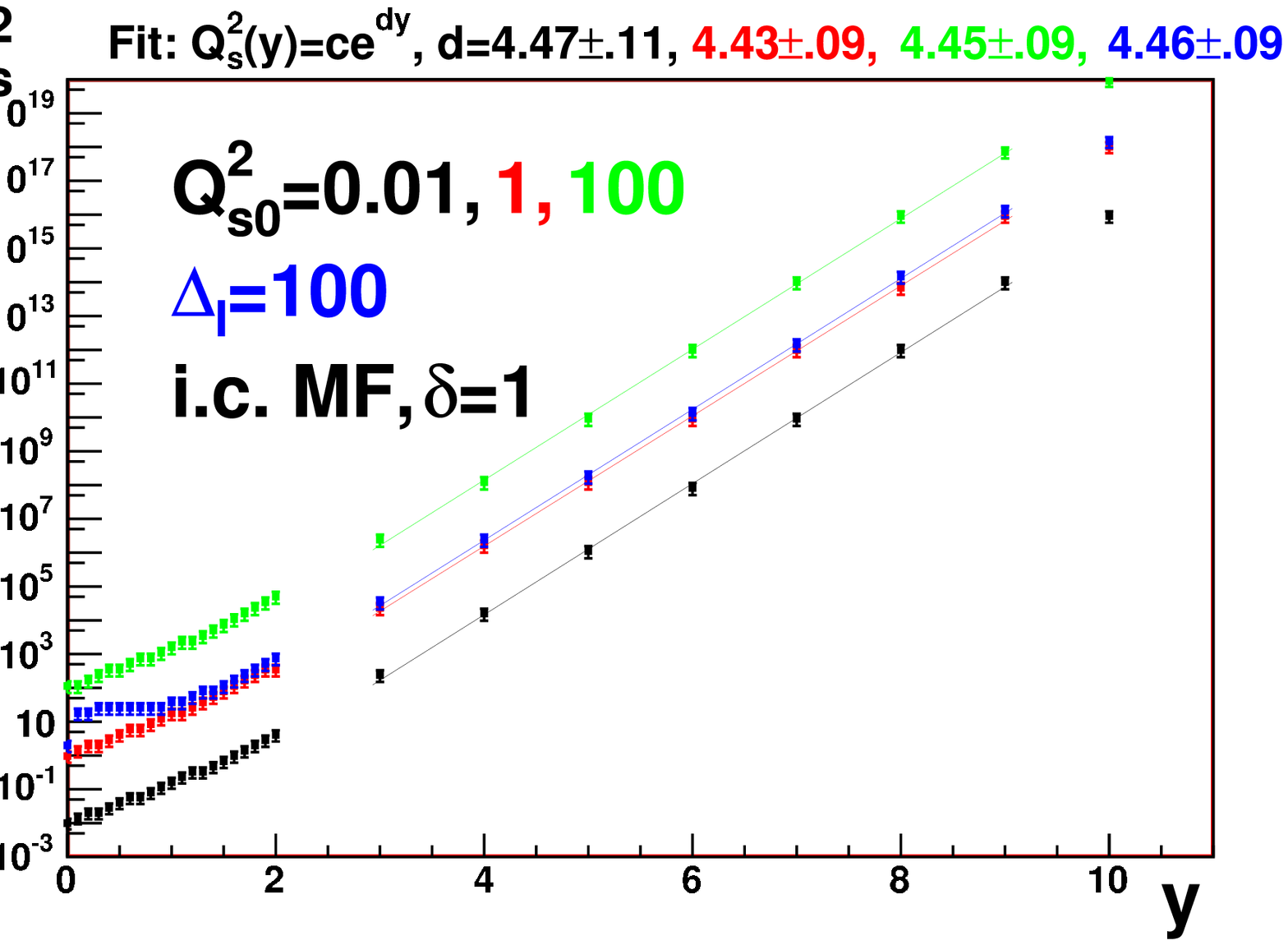,height=7.5cm,angle=90}
\vskip 0.3cm
\epsfig{file=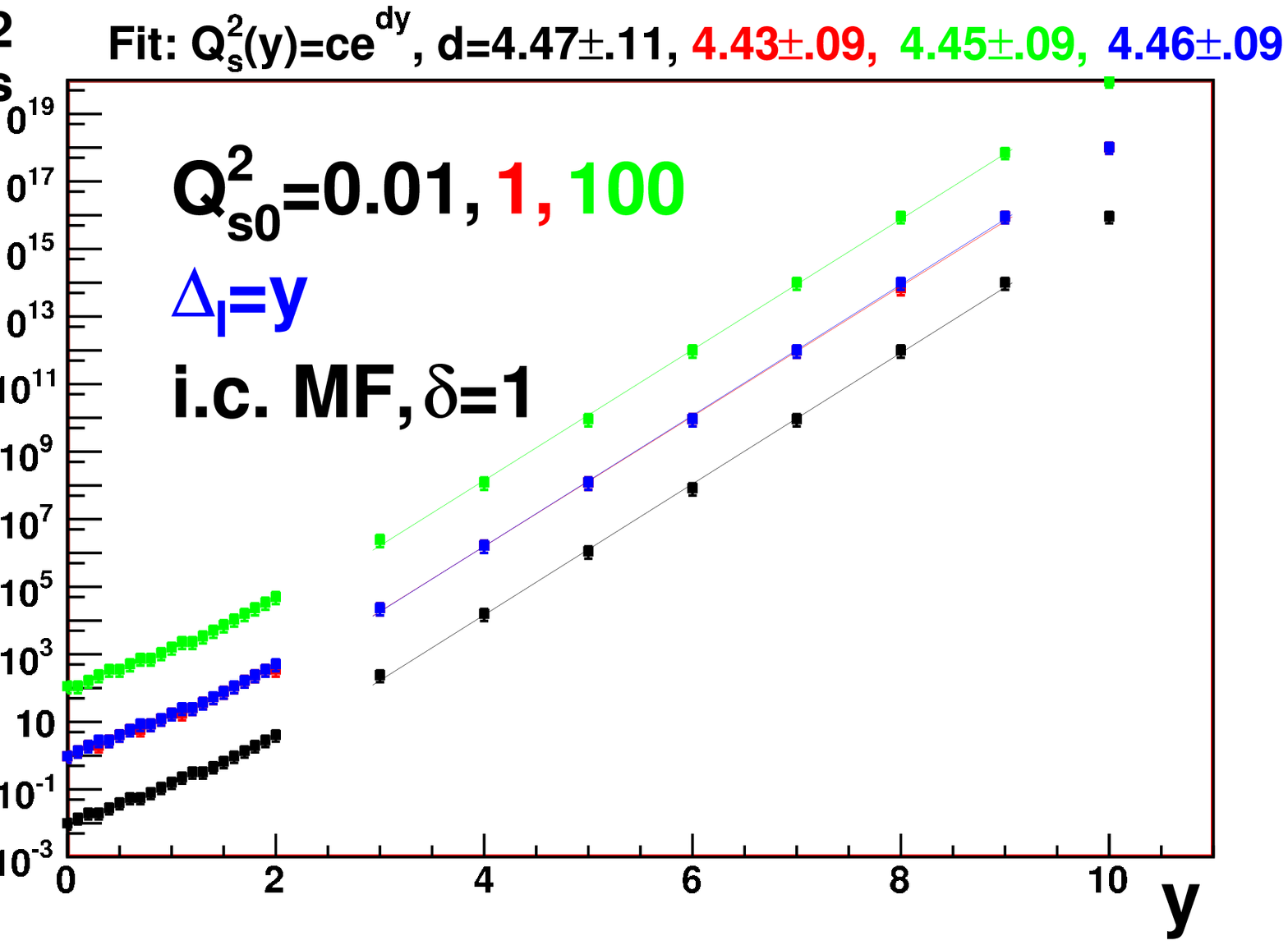,height=7.5cm,angle=90}\hskip 0.5cm\epsfig{file=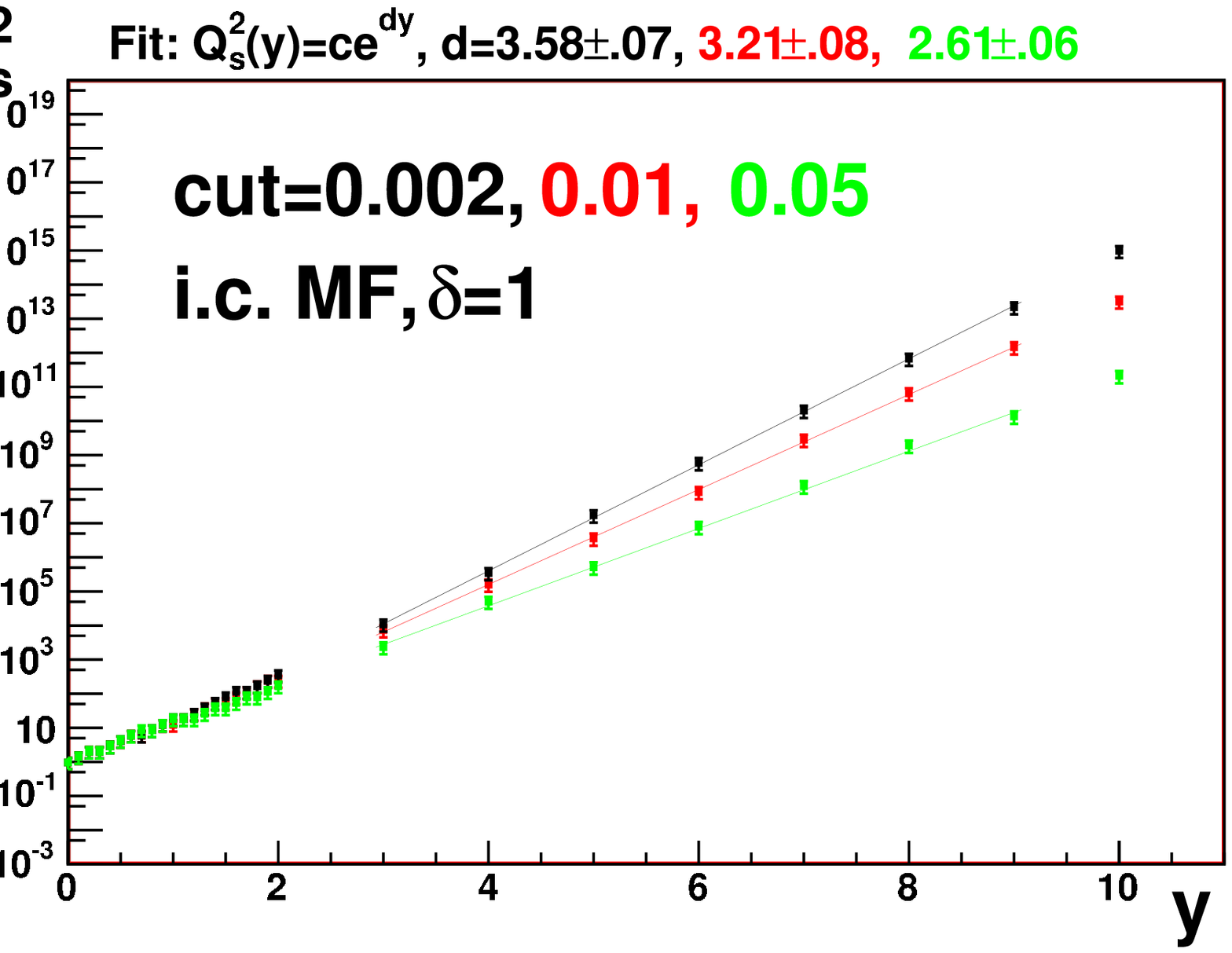,height=7.5cm,angle=90}
\end{center}
\vskip -1.cm
\caption{$Q_s^2$ versus $y$ for evolution starting
from MF initial conditions. In the upper and lower-left
plots, the results for the evolution of individual configurations are shown
for $Q_{s0}^2=0.01$, 1 and 100 in black, red and green respectively, while in
blue the results for linear ($\Delta=100$, upper-left plot) and logarithmic
($\Delta_l=100$, upper-right plot, and $\Delta_l=y$, lower-left plot)
are presented. In the lower-right plot, 
results of the cut-off evolution for $\kappa=0.002$ (black), 0.01 (red) and
0.05 (green) are presented. In the plots, straight lines correspond to fits to
$Q_s^2(y)=ce^{dy}$ in the region $3<y<9$, with the values of $d$
indicated in the plots.}
\label{fig5}
\end{figure}

To study the rapidity evolution of the saturation scale, we determine it as
indicated in the previous Subsection.  An error, corresponding to the distance
between neighbors in the grid at the saturation scale, is assigned to every
value of the saturation scale.  In Fig.~\ref{fig5} we show the results for the
evolution of the saturation scale with reduced rapidity $y$. In order to
quantify eventual differences, we have performed a fit to an exponential
form\footnote{The value of $d$
expected~\cite{Iancu:2002tr,Mueller:2002zm,Munier:2003vc} is $d\simeq 4.88$;
for evolution of individual configurations and for the average, we get a
smaller value, $d\simeq 4.45$, due to the neglected sub-leading
terms~\cite{Munier:2004xu,Golec-Biernat:2004sx,Albacete:2004gw,Enberg:2005cb}
in the $y$-behavior of the saturation scale at these large but finite values
of $y$.} $Q_s^2(y)=ce^{dy}$ in the region
$3<y<9$.

The averaging procedure shows no sizable effect, a fact which contradicts the
expectations from sFKPP but can be easily
understood as BK evolution is
known~\cite{Armesto:2001fa,Lublinsky:2001bc,Albacete:2003iq,
Iancu:2002tr,Munier:2003vc} to lead asymptotically to a universal wave front
moving with universal velocity (for `supercritical' initial conditions, see
the footnote in Subsection~\ref{initial}). Thus, asymptotically every
individual configuration (even those arising from initial conditions not only
with different initial saturation scale but also with different functional
shape) will move with the same velocity, and the average will be characterized
by this common velocity.

On the other hand, the cut-off evolution shows a strong influence on the
$y$-evolution of the saturation scale in agreement with expectations from
sFKPP: the larger the value of the cut-off parameter $\kappa$, the slower the
evolution\footnote{In the framework of sFKPP, behaviors proportional to
$(\ln{\kappa})^{-1}$ and $\kappa^{-1}$ are expected in the limit of
weak~\cite{Iancu:2004es,Enberg:2005cb,Brunet:2005bz} and
strong~\cite{Soyez:2005ha,Marquet:2005ak} noise respectively. Our results show
a mild behavior $\propto \kappa^{-0.095}$.}. This corresponds to the fact that
the evolution is driven~\cite{Albacete:2003iq, Iancu:2002tr,Munier:2003vc} by
the large-$k$ regions, as discussed previously in Subsection~\ref{scaling}.

Other interesting quantity which we can examine\footnote{We thank Alex Kovner
for drawing our attention on this point.} is not the saturation scale of the
averaged solution, but the average saturation scale. In the language
of~\cite{Munier:2003vc,Munier:2003sj,Munier:2004xu}, the evolution of the
saturation scale with rapidity shows the speed of the wave front in its
movement towards higher transverse momenta. Thus the dispersion of the
saturation scale at a given rapidity will reflect the spread of the ensemble
of wave fronts coming through evolution from a given ensemble of initial
conditions. It is then a very good quantity to characterize the effect of the
evolution on a possible transition from geometric to diffusive
scaling~\cite{Hatta:2006hs}.

\begin{figure}[!ht]
\begin{center}
\epsfig{file=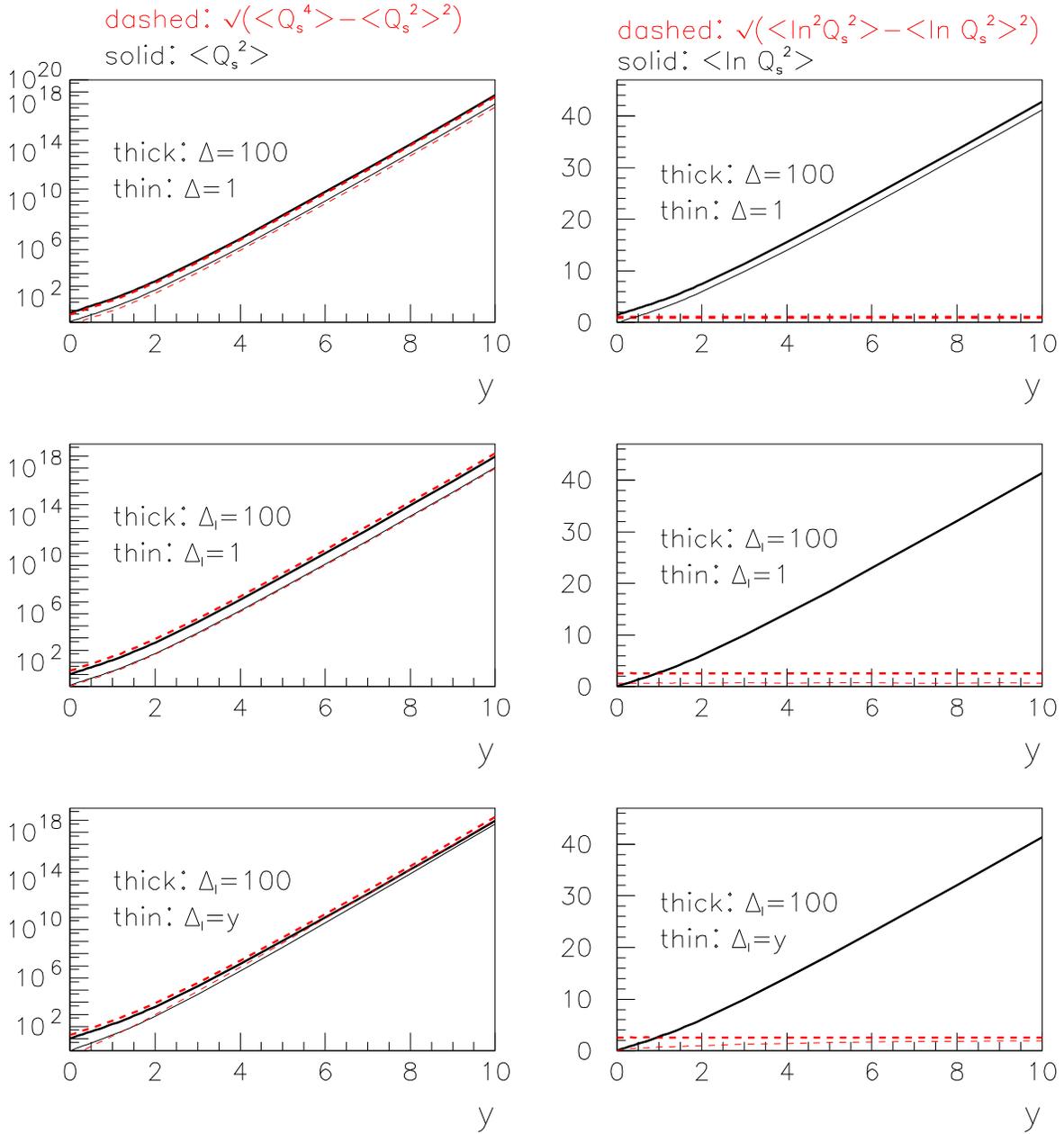,width=15.5cm}
\end{center}
\vskip -1.cm
\caption{$\langle Q_s^2(y)\rangle$ (black solid lines in the plots on the
left),
$\langle \ln Q_s^2(y)\rangle$ (black solid lines in the plots on the right),
and their corresponding dispersions (red dashed lines in the plots on the left
and right respectively) versus $y$,
for linear (upper plots) and Gaussian (middle and lower plots)
averaging
with $\Delta,\Delta_l=100$ (thick lines), and $\Delta=1$,
$\Delta_l=1$
and $\Delta_l=y$ (thin lines in the
upper, middle and lower plots respectively),
for evolution starting
from MF, $\delta=1$, initial conditions. The difference between thick and thin
red dashed lines in the upper-right plot, and between thick and thin black
solid lines in the middle- and lower-right plots, is numerically very small
and hardly visible.}
\label{fig6}
\end{figure}

We define the average of any observable $\langle {\cal O}(y)\rangle $ by
substituting $\phi(k)$ by ${\cal O}(y)$ in Eq.~(\ref{eq5}). In Fig.~\ref{fig6}
we plot the average values $\langle Q_s^2(y)\rangle$ and $\langle \ln
Q_s^2(y)\rangle$ and the corresponding dispersions, versus rapidity, for
linear and logarithmic weight functions, for two different fixed values of
$\Delta, \Delta_l$, and also for $\Delta_l=y$.  In the case of fixed $\Delta,
\Delta_l$ both the average $Q_s^2(y)$ and its dispersion show the same
dependence with $y$, so the width of the ensemble of wave fronts gets wider
and wider with increasing rapidity. On the other hand, the dispersion of $\ln
Q_s^2(y)$ stays constant (within our numerical accuracy) during evolution - a
behavior opposite to the linear $y$-dependence found in
sFKPP~\cite{Iancu:2004es,Enberg:2005cb,Brunet:2005bz,Soyez:2005ha,Marquet:2005ak}.
Besides, smaller initial widths result in smaller widths after evolution. We
expect these characteristics to remain true for large enough rapidity even if
the weight functions for averaging are more involved, or even for initial
ensembles containing different shapes. This is due to the fact that BK
evolution eventually leads to a universal shape which, once roughly built,
will evolve with the characteristics we have discussed.  On the other hand,
for $\Delta_l=y$ the dispersion of $\ln Q_s^2(y)$ increases with increasing
rapidity roughly linearly in agreement with expectations from sFKPP, though
with a coefficient smaller than expected from the dispersion introduced in the
weight function.

\section{Discussion} \label{disc}

Much theoretical effort has been directed recently to go beyond the standard
JIMWLK framework to study the small-$x$ structure of hadrons: the role of
correlations~\cite{Levin:2003nc,Janik:2004pg,Levin:2004yd,Iancu:2004es,Kovner:2005jc,Marquet:2005ak},
of discreteness in gluon
emissions~\cite{Salam:1995uy,Iancu:2004es,Kharzeev:2005gn,Marquet:2005er} and
the relevance of Pomeron
loops~\cite{Mueller:2004se,Iancu:2004iy,Mueller:2005ut,Levin:2005au,
Iancu:2005nj,Blaizot:2005vf,Marquet:2005hu,Kovner:2005nq,
Kovner:2005en,
Kovner:2005uw,Hatta:2005ia,Hatta:2005rn,Shoshi:2005pf,Iancu:2005dx}.
Numerical studies are just
starting~\cite{Rummukainen:2003ns,Soyez:2005ha,Enberg:2005cb,Abramovsky:2005vm,
Marquet:2005er}.

In this paper we contribute to this subject through an attempt to quantify
numerically the relevance of two of these aspects, namely correlations through
averaging~\cite{Kovner:2005jc} and discreteness of gluon
emissions~\cite{Iancu:2004es}, on standard BK evolution. The averaging of
individual configurations results in scaling violations, but their size around
the saturation scale is not much larger than the violations already present
in standard evolution at a large but finite rapidity.  On the other hand, the
evolution of the saturation scale with rapidity, and the large-$k$ behavior of
the solutions, seem to be hardly affected. Finally, the dispersion in the
saturation scale of the individual solutions increases with rapidity at least
as fast as the corresponding average, indicating that the width of the
ensemble of evolved solutions gets wider and wider with increasing rapidity.

The implementation of discreteness in the evolution through a cut-off does not
induce any further scaling violation. On the other hand, the saturation scale
evolves slower with rapidity with increasing value of the cut-off, and the
solutions at large $k$ become steeper than in the standard BK evolution.

When compared with the expectations from the sFKPP
equation~\cite{Iancu:2004iy,Iancu:2005nj,Soyez:2005ha,Enberg:2005cb,Brunet:2005bz,Marquet:2005ak},
we conclude than none of the proposed modifications is able to reproduce all
the features found there. The implementation of a cut-off slows down the
evolution in accordance with expectations, but results in a better
fulfillment of scaling, at variance with sFKPP. The averaging procedure, when
implemented through a dispersion increasing with rapidity, leads to 'diffusive'
wave fronts~\cite{Hatta:2006hs}, but the rapidity evolution of the saturation
scale is not affected, contrary to expectations.

In view of the differences among the effects of different modifications on
standard BK evolution and with the results of sFKPP, of the fact that the
experimentally accessible regions of rapidity are pre-asymptotic and dominated
by the initial
conditions~\cite{Armesto:2004ud,Albacete:2004gw,Albacete:2005ef}, and of the
additional uncertainties on the form of the averaging weight function and/or
size of the cut-off, any claim on phenomenological implications of our study
looks premature.  Besides, our implementation of these modifications of BK
evolution is simplistic, so their results could only be taken as hints of the
possible effects of a proper application of such modifications. Nevertheless,
we find their effects large and diverse enough to justify the further
investigations, both on the analytical and on the numerical side, currently
under development to rigorously develop and implement all these new ideas.
Work in this direction is in progress.

\vskip 0.8cm
\noindent{\bf Acknowledgments:}
We thank M.~A.~Braun, J.~Jalilian-Marian,
A.~H.~Mue\-ller, K.~Tuchin and U.~A.~Wiedemann for useful discussions.
Special thanks are given to J.~L.~Albacete, E.~Iancu,
A.~Kovner and M.~Lublinsky for
many fruitful discussions and suggestions,
and a critical reading of the manuscript.
N.A. acknowledges financial support of Ministerio de Educaci\'on y
Ciencia
of Spain under a contract Ram\'on y Cajal, and of
CICYT of Spain under project
FPA2002-01161. J.G.M. acknowledges the partial financial support from the Funda\c c\~ao para
a Ci\^encia e a Tecnologia of Portugal under contract
SFRH/BPD/12112/2003. N.A. also thanks IST, and J.G.M.
Departamento de F\'{\i}sica
de Part\'{\i}culas at the Universidade de Santiago de Compostela, for
warm hospitality while part of this work was done.

\end{document}